\documentclass[aps,prl,a4paper,amsmath,amssymb,amsfonts,twocolumn,
floatfix,superscriptaddress,showkeys]{revtex4-1}
\usepackage{bm,amsmath,amssymb,amsfonts}
\usepackage[dvips]{graphicx}
\usepackage[utf8]{inputenc}
\usepackage[normalem]{ulem}
\usepackage{color}
\usepackage{xcolor}

\definecolor{dgreen}{rgb}{0,0.6,0}
\usepackage{hyperref}

\hypersetup{
    pdfnewwindow=true,     
    colorlinks=true,       
    linkcolor=red,     
    citecolor=blue,        
    filecolor=blue,        
    urlcolor=blue          
}
\begin{document}
\title{Observation of complete delocalization in disordered photonic lattices}
\author{Biplab Pal}
\email[E-mail: ]{biplab@nagalanduniversity.ac.in}
\affiliation{Department of Physics, School of Sciences, 
Nagaland University, Lumami 798627, Nagaland, India}
\author{Rodrigo A. Vicencio}
\email[E-mail: ]{rvicencio@uchile.cl}
\affiliation{Departamento de F\'{i}sica, Facultad de Ciencias F\'{i}sicas y Matem\'{a}ticas, 
Universidad de Chile, Santiago, Chile} 
\affiliation{Millennium Institute for Research in Optics - MIRO, Santiago, Chile} 
\date{\today}
\begin{abstract}
We present the exceptional phenomenon of complete absence of Anderson localization, and perfect transmission of 
particles, in a completely disordered diamond-dot chain. We analytically show a proof for the condition to observe 
this exceptional phenomenon, based on a transparent window emerging from a geometrical condition. We support our 
theoretical prediction by numerical simulations and direct experimental observation of the transmission probabilities 
of the light in a femtosecond laser-written diamond-dot photonic lattices. We additionally show that for a $\pi$ 
effective magnetic flux, extreme localization of the light in the same system may occur, independently on the 
specific geometry. Our results open up an excellent platform for controlling the transmission of energy from 
ballistic to zero transmission, in a completely disordered lattice system.
\end{abstract}
\maketitle
%

The transport properties of a quantum system are controlled by the nature of their eigenstates. For a perfectly ordered 
quantum array, described by a lattice model, all the single-particle quantum states are, in general, perfectly extended 
Bloch modes~\cite{Bloch-1929}, determining a ballistic transmission of the particles through the entire system. On the 
other hand, in a completely disordered quantum system, the states become localized in the presence of any amount of 
disorder. This leads to the complete absence of diffusion of waves, known as the Anderson localization 
(AL)~\cite{Anderson-prb-1958,Ramakrishnan-prl-1979}. This phenomenon has been a long-standing area of research interest 
in the condensed-matter physics community over many decades; however, only about two decades ago, the direct experimental 
observation of Anderson localization of light~\cite{schwartz_transport_2007,Lahini08} and matter~\cite{Billy-Nature-2008,
Roati-Nature-2008} waves was achieved. In contrast to AL, compact localization of all single-particle states is also 
possible in a completely disorder-free linear system. This originates purely due to the lattice geometry and the insertion 
of effective magnetic fluxes, a phenomenon known as the Aharonov-Bohm (AB) caging~\cite{Vidal-prl-1998,Vidal-prl-2000,
Sebabrata-prl-2018,Rodrigo-prl-2022}.

Some examples show exceptions to the conventional scenario of AL in disordered systems, within the framework of tight-binding 
formalism. These examples are attributed to the different kinds of correlations between the Hamiltonian parameters; e.g., 
certain special positional correlation between the on-site potentials of a 1D chain known as the random-dimer 
model~\cite{Dunlap-prl-1990,Naether_2013}, correlations between the diagonal and off-diagonal matrix elements of the Hamiltonian 
in any spatial dimension~\cite{Dunlap-prb-1989}, a 1D model with the on-site energies exhibiting long-range correlated 
disorder~\cite{de-Moura-prl-1998}, multi-leg ladder models with correlated disorder~\cite{Sedrakyan-prb-2004,Sedrakyan-pra-2011}, 
quasiperiodicity and interactions~\cite{Flach12}, to name a few. However, in all these models, the localization-delocalization 
transition and the appearance of the extended states happened only at a special discrete set of energy eigenvalues. Later on, the 
possibility of engineering the bands of extended states for the whole allowed range of energies, in a class of completely disordered 
and quasi-periodically ordered lattice models, was also proposed theoretically~\cite{Pal-epl-2013,Pal-pla-2014,Pal-physe-2014}. 
However, to the best of our knowledge, the experimental observation of such a complete absence of AL for the whole allowed range 
of band energies has not been reported previously. 
\begin{figure}[t!]
\centering
\includegraphics[clip, width=\columnwidth]{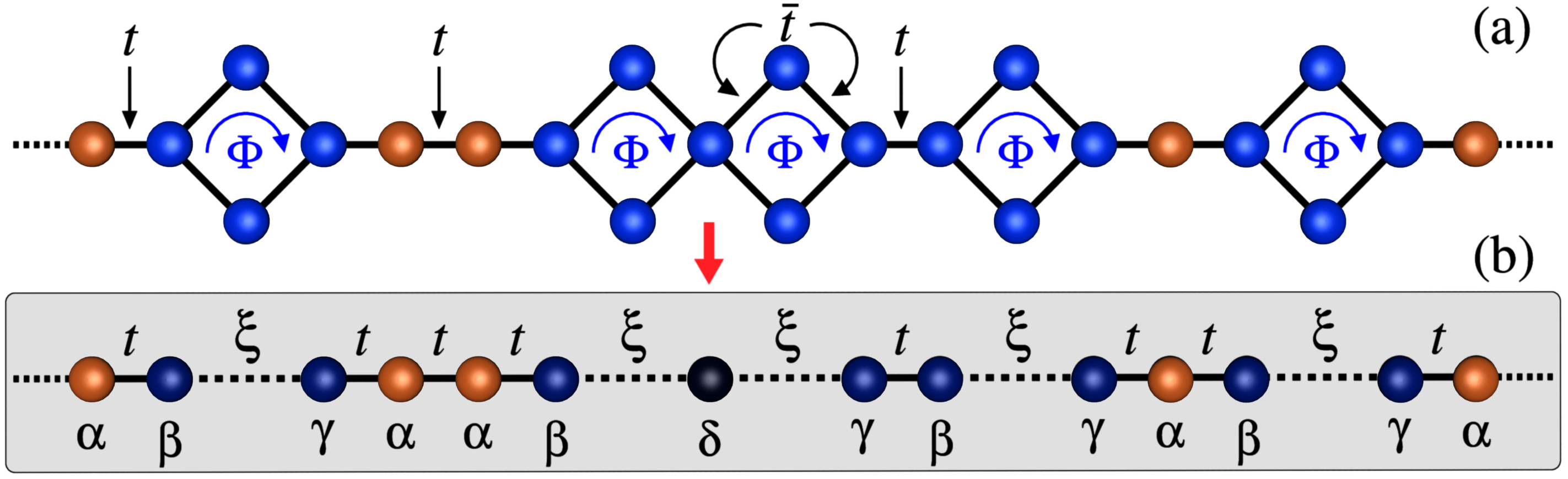}
\caption{(a) Diagram of a random diamond-dot chain. $t$ and $\bar{t}$ 
define the respective couplings, and $\Phi$ indicates the magnetic flux. 
(b) The renormalized version of the original DD chain, with $\xi$ representing 
the effective horizontal coupling in each renormalized diamond plaquette.}
\label{fig:Lattice-model}
\end{figure}

In this letter, we report the direct experimental observation of the complete absence of AL and a perfectly ballistic 
transmission of light in completely disordered diamond-dot (DD) photonic lattices. The experiments are implemented in 
lattices fabricated by means of the femtosecond (fs) laser-writing technique~\cite{szameit_discrete_2005,Rodrigo-prl-2024}, 
allowing us to completely design every lattice realization. We analytically find the specific geometrical conditions 
that characterize the DD model and predict a perfect transmission probability of the particles, on completely disordered 
random lattices, for the whole allowed range of energies. This analytical proof is extensively supported by the experimental 
data, which exhibit the evidence of the appearance of this exceptional phenomenon of complete absence of AL. In addition 
to this, we have experimentally demonstrated that this photonic lattice also supports the AB caging phenomenon, which 
produces null transmission for an effective $\pi$ magnetic flux. 

We consider the tight-binding model of a disordered DD chain as shown in Fig.~\ref{fig:Lattice-model}(a). One can easily 
map the original DD chain into an effective 1D model, as depicted in Fig.~\ref{fig:Lattice-model}(b), using a standard 
real-space renormalization technique~\cite{Pal-epl-2013,Pal-pla-2014,Pal-physe-2014}. A 1D tight-binding lattice can be 
described by the following discretized version of the Schr\"{o}dinger equation: 
%
\begin{equation}
\left(E-\varepsilon_{n}\right)\psi_{n} = t_{n,n+1}\psi_{n+1} + t_{n,n-1}\psi_{n-1}\ ,
\label{eq:difference-eqn} 
\end{equation}
%
where $\varepsilon_{n}$ is the on-site potential at the $n$-th site, $t_{n,n \pm 1}$ are the nearest-neighbor hopping 
amplitudes (couplings), and $\psi_{n}$ is the wavefunction amplitude at the $n$-th site. Using Eq.~\eqref{eq:difference-eqn}, 
we can find out the relation between the wavefunction amplitudes of the neighboring sites in a 1D chain through the transfer 
matrix. We define $\Phi$ as the magnetic flux in each diamond plaquette, that is effectively inserted through the specific 
hoppings (see Supplemental Material for more details~\cite{SM}). We set $\varepsilon_{n}=\varepsilon_{0}$ for all the sites. 
\begin{figure}[t!]
\centering
\includegraphics[clip, width=\columnwidth]{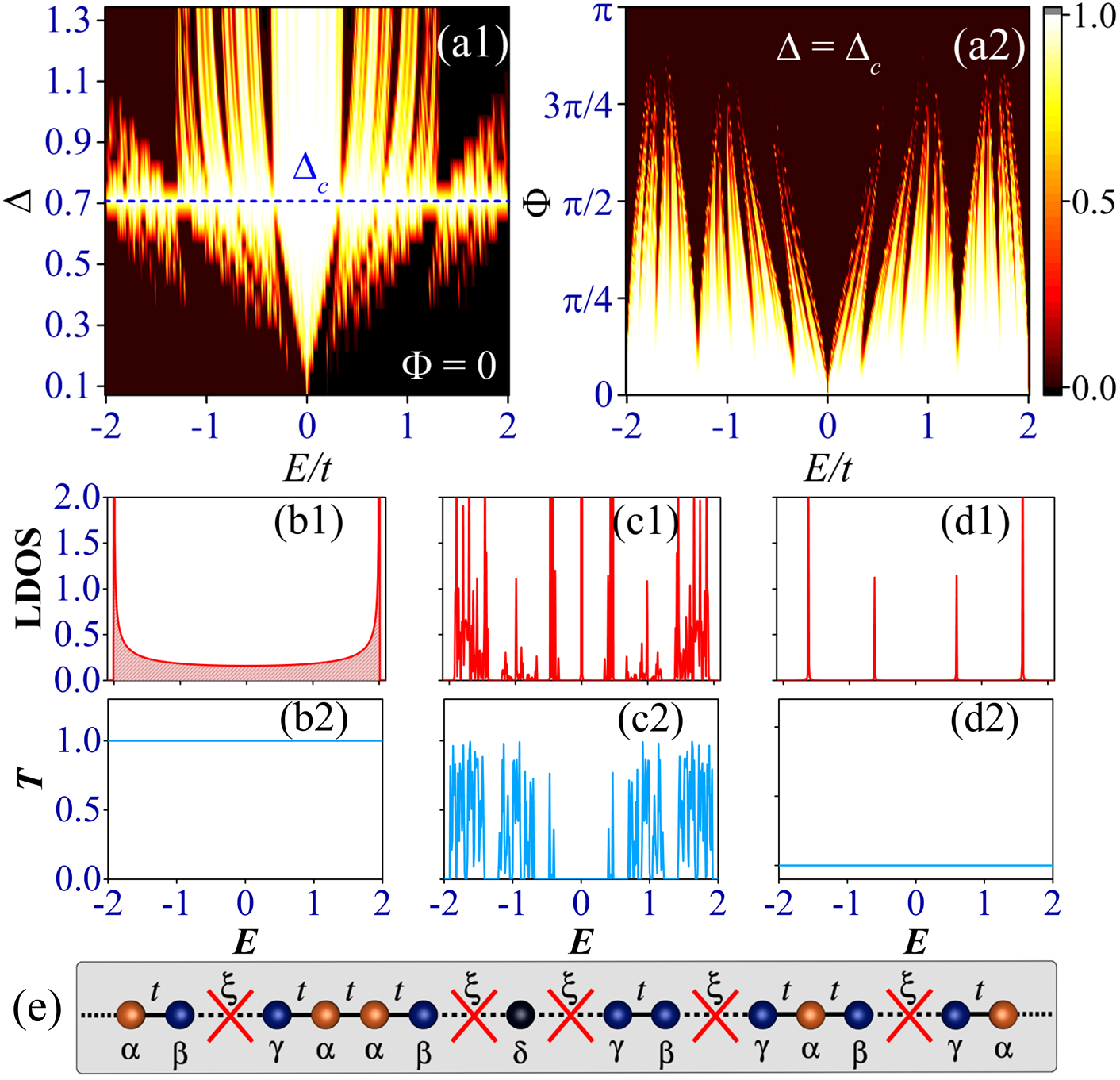}
\caption{(a1) Density plot of $T$ as a function of $E$ and $\Delta$, for 
$\Phi=0$. (a2) same as (a1) as a function of $E$ and $\Phi$, for 
$\Delta_{c} = 1/\sqrt{2}$. (b1) and (b2) LDOS and transmission probability, 
respectively, under the \emph{resonance condition} $\Delta_{c}$ and $\Phi=0$. 
(c) and (d) are the same as (b), but under an \emph{off-resonance condition} 
at flux $\Phi=\pi/2$ and $\Phi=\pi$, respectively. (e) Lattice diagram for 
$\Phi=\pi$, with the effective hopping $\xi=0$ leading to the AB caging.}
\label{fig:Transmission}
\end{figure}

From Fig.~\ref{fig:Lattice-model}(a), we define the hopping ratio $\Delta\equiv \bar{t}/t$, as the critical geometrical 
parameter of this problem. One can easily work out and identify that, all the involved transfer matrices for the system 
and their combinations commute with each other, when we set the condition $\Delta = \Delta_{c}\equiv 1/\sqrt{2}$ at 
$\Phi=0$~\cite{SM}. We emphasize upon the fact that, this happens \emph{independently of the energy $E$ of the particle}; 
i.e., for all the allowed energy eigenvalues of the system. That means that under this special condition, and for $\Phi=0$, 
the completely disordered DD chain effectively turns into a perfectly ordered 1D lattice, giving us the absolutely continuous 
energy spectrum with all the states being \emph{perfectly extended} Bloch modes. We quantify this analytical exact proof in 
Fig.~\ref{fig:Transmission}(a1), where we show the two-terminal transmission probability ($T$) of the particle through a 
disordered DD chain as a function the energy $E$ and the ratio $\Delta$. From Fig.~\ref{fig:Transmission}(a1), it is clearly 
visible that, only around the critical value $\Delta_{c}$, we get a ballistic transmission through the system, irrespective 
of the energy of the particle. It is noticeable that, as we deviate from $\Delta_{c}$, we observe that the transmission 
probability through the system is drastically reduced (see darker colors). 

On the other hand, once the condition $\Delta = \Delta_{c}$ is set, it will be possible to detune the \emph{resonance condition} 
by tuning an external magnetic flux $\Phi$ to a nonzero value [see Fig.~\ref{fig:Transmission}(a2)]. We notice how the 
transmission spectrum splits into narrower bands while increasing $\Phi$, observing zero transport at $\Phi=\pi$. 
Figs.~\ref{fig:Transmission}(b)-(d) show the local density of states (LDOS) and $T$ versus $E$, both under the \emph{resonance} 
and \emph{off-resonance} conditions. We observe how for $\Phi\neq 0$, the LDOS diagrams filament into narrower allowed energy 
regions, with $T\neq 0$ only at very specific $E$ values. At flux $\Phi = \pi$, all the states are extremely localized with a zero 
transmission probability, as the effective hopping $\xi = 2\bar{t}^{2}\cos(\Phi/2)/(E-\varepsilon_{0})$ becomes exactly zero 
[see Fig.~\ref{fig:Transmission}(e)]. This is a consequence of the AB caging phenomenon and the emergence of an all-flat band 
regime~\cite{Rodrigo-prl-2022,Yang2024}, with only compact (zero tail) eigenstates. 

\begin{figure}[t!]
\centering
\includegraphics[clip, width=\columnwidth]{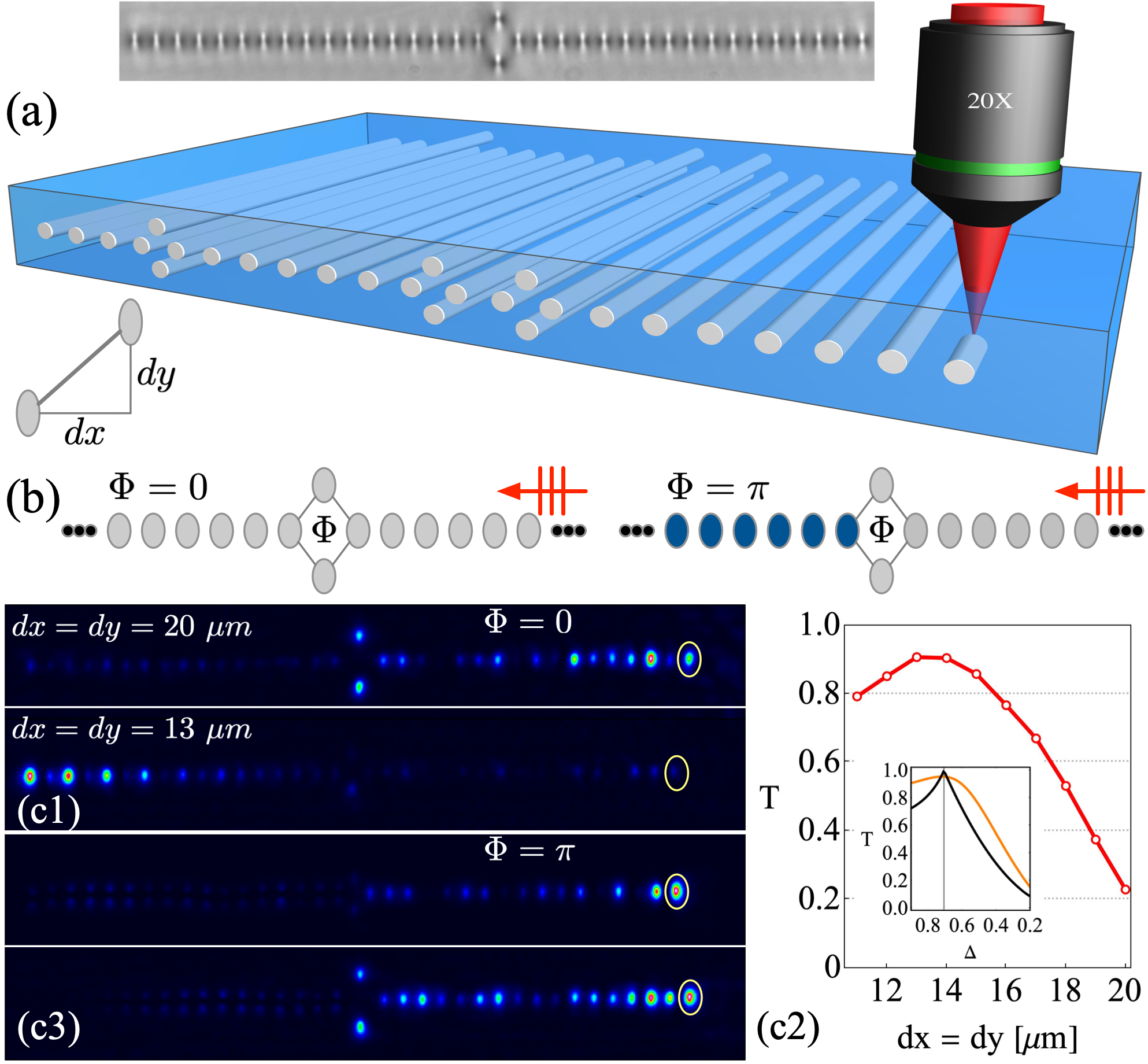}
\caption{(a) Sketch of the fs laser-writing technique. Upper inset: Microscope 
image of a lattice with a single diamond plaquette. Bottom inset: diamond geometry. 
(b) Sketch of a single-diamond lattice with zero (left) and $\pi$ (right) fluxes 
(gray and blue colors correspond to $S$ and $P$ sites, respectively). 
(c1) Intensity output profiles for $\Phi=0$, after a right edge excitation, and 
or $dx=20$ $\mu$m (top) and $13$ $\mu$m (bottom). (c2) Transmission data $T$ vs.\ 
diamond geometry $dx=dy$ for an edge excitation with $\Phi=0$. Inset: Numerical 
$T$ (orange) and theoretical $\left<T\right>_k$ (black) vs.\ $\Delta$. (c3) Same 
as (c1) for $\Phi=\pi$.} 
\label{fig3}
\end{figure}

Now, we implement the DD model on a photonic configuration. We consider lattices composed of a set of optical waveguides, which 
are very well described by tight-binding-like models~\cite{lederer_discrete_2008}. First of all, we explore the simplest system 
consisting of a single diamond inserted on a 1D lattice [see Fig.~\ref{fig3}(a)-top]. We start by computing the plane wave 
transmission $T$ across a single diamond plaquette~\cite{SM}, obtaining 
%
\begin{equation}
T(k,\Delta)=\frac{4\Delta^4 \sin^2 k}{[(1-2\Delta^2)^2\cos^2 k+4\Delta^4 \sin^2 k]}\ ,
\label{Tksingle} 
\end{equation}
%
with $k$ being the horizontal quasi-momentum. Here, we immediately notice that any plane wave (i.e., any $k$ or any $E$) has full 
transmission $T=1$ at $\Delta = 1/\sqrt{2}$. In other words, under the critical condition $\Delta=\Delta_{c}$, the system becomes 
transparent to any propagating wave, and the diamond plaquette transforms into an effective single site inserted on a 1D lattice.

The first experiment we perform consists on studying the transmission response of this configuration, such as to find the geometrical 
parameters necessary to observe the critical phenomenology. The coupling (hopping) among waveguides decays exponentially over the 
inter-site distance~\cite{szameit_discrete_2005}, therefore the calibration of the diamond plaquette geometry is mandatory. We 
fabricate several lattices using the fs laser-writing technique~\cite{szameit_discrete_2005}, as it is sketched in Fig.~\ref{fig3}(a). 
For simplicity, we consider only symmetric geometries $dx=dy$ for the plaquettes [see the inset in Fig.~\ref{fig3}(a)]. We implement 
this experiment by considering an effective magnetic flux of $0$ and $\pi$, as sketched in Fig.~\ref{fig3}(b). A zero flux demands 
a lattice having only single-mode waveguides (gray ellipses), while flux $\Phi=\pi$ requires the insertion of dipolar $P$ (darker 
ellipses) waveguides~\cite{guzman-silva_experimental_2021,Rodrigo-prl-2022,MultiorbitalVic}. We fabricate different diamond geometries 
such to approach the critical condition. This process is not trivial, as the reduction of the plaquette size implies an increment 
of second-order couplings, which could start affecting the dynamics. Fig.~\ref{fig3}(c1) shows intensity output profiles for $\Phi=0$ 
and for the indicated geometries. We clearly notice the effect of increasing the diagonal coupling $\bar{t}$ while decreasing the 
diamond dimensions from $dx=20$ to $13\ \mu$m. We observe almost no transmission for $dx=20\ \mu$m ($\Delta<\Delta_c$), while a 
completely transparent regime for $dx=13\ \mu$m ($\Delta\sim\Delta_c$). Fig.~\ref{fig3}(c2) compiles our experimental results for 
different plaquette geometries and for an excitation wavelength of $\lambda=790$ nm. We observe a clear maximum transmission at 
around $dx=dx_c=13\ \mu$m with $T\sim 91\%$, indicating that the condition $\Delta = \Delta_{c}$ is satisfied around this geometry. 

The orange line in Fig.~\ref{fig3}(c2)-inset presents the results obtained by numerically integrating the dynamical version of the 
model in Eq.~\eqref{eq:difference-eqn}, after exciting the lattice edge and measuring the transmitted energy. We observe a quite 
similar curve compared to the experimental data (red circles), as a confirmation of the critical plaquette geometry around $dx_c$. 
The averaged plane wave transmission $\left<T\right>_k$, obtained after averaging $T$ from Eq.~\eqref{Tksingle} in the interval 
$k\in\{0,\pi/2\}$, is shown in black in the Fig.~\ref{fig3}(c2)-inset. $\left<T\right>_k$ gives a narrower distribution around the 
critical value $\Delta_{c}$, as Eq.~\eqref{Tksingle} has a different response over $k$. Both cases show that a perfect transmission 
is obtained around the critical ratio $\Delta_{c}$, indicated by a vertical line. At flux $\Phi=\pi$, any geometry may produce zero 
transmission as the light is not able to pass through the diamond plaquette due to the AB caging effect~\cite{Vidal-prl-2000}: the 
light transiting through the upper ``channel" is out of phase with respect to the one at the bottom channel, therefore destructively 
interfering when exiting. Fig.~\ref{fig3}(c3) shows that for $\Phi=\pi$, and two different geometries, only a negligible amount of 
energy passes through the plaquette, with $T\sim 1\%$~\cite{SM}. 
\begin{figure}[t!]
\centering
\includegraphics[clip, width=\columnwidth]{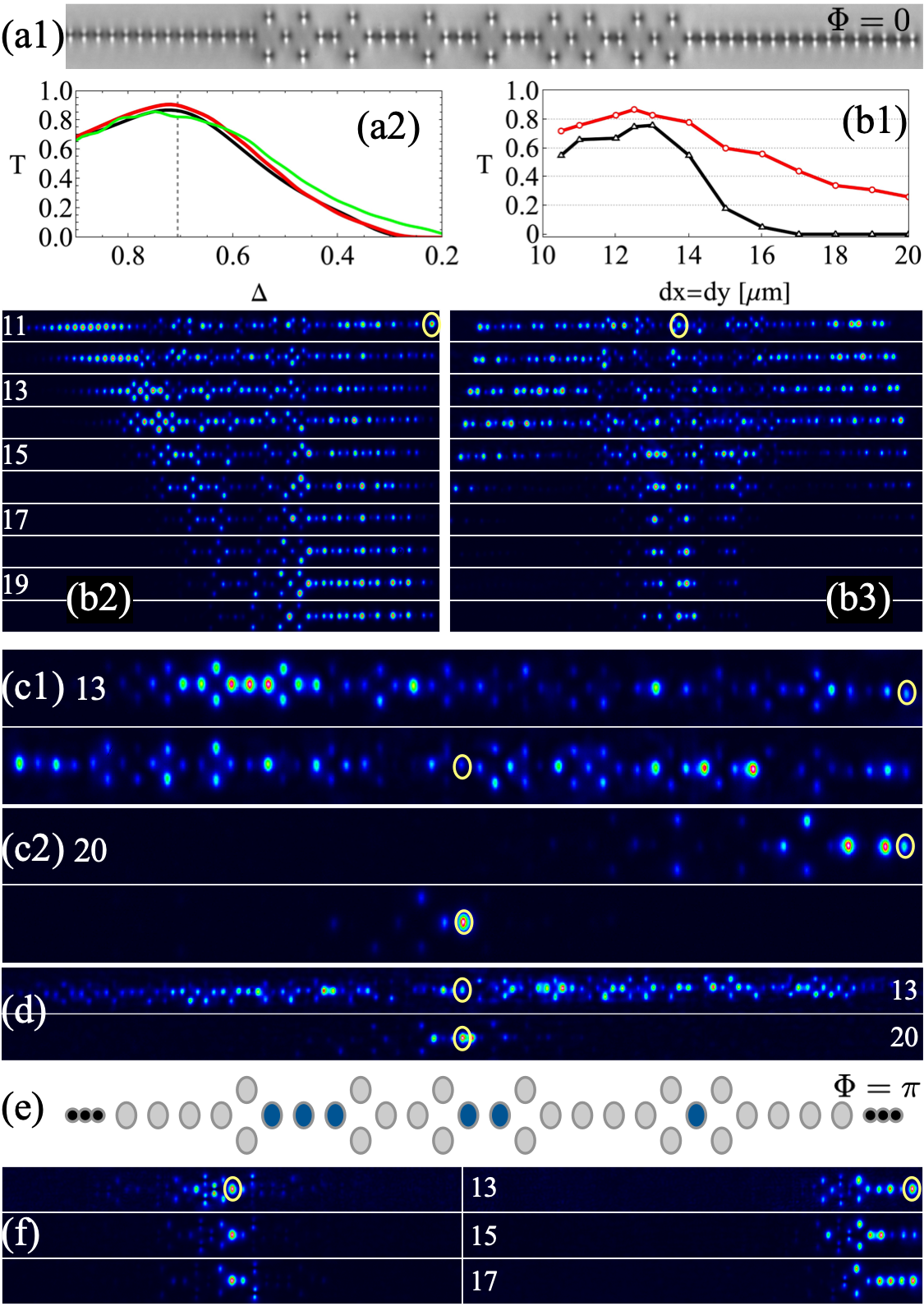}
\caption{(a1) Microscope image of a random realization with 9 plaquettes for 
$\Phi=0$. (a2) Numerical transmission versus $\Delta$ for a lattice with 9 
plaquettes, for a single (red) and averaged (black) edge and bulk (green) 
excitations. (b1) Extracted transmission data for the intensity output images 
(IOI) shown in (b2) and (b3), for $dx\in\{11,20\}\ \mu$m. (c1) and (c2) IOI 
for a lattice with $16$ diamond plaquettes for $dx=13$ and $20$ $\mu$m, 
respectively, for edge (top) and bulk (bottom) excitations. (d) IOI for a 
bulk excitation of a lattice with $32$ plaquettes for $dx=13$ (top) and 
$20$ (bottom) $\mu$m. (e) Sketch for $\Phi=\pi$. (f) IOI for lattices with 
$16$ plaquettes and $\Phi=\pi$, for $dx=13$ $\mu$m (top), $15$ $\mu$m (center), 
and $17$ $\mu$m (bottom), for a bulk (left) and edge (right) excitations. 
Yellow ellipses indicate the excitation positions for $\lambda=750$ nm.} 
\label{fig4}
\end{figure}

From Fig.~\ref{fig:Transmission}(a) we know that a random distribution of plaquettes may induce Anderson-like localization away 
from the transparent window $\Delta = \Delta_{c}$. The light beam will interact with randomly distributed plaquettes that will 
generate random transmitted and reflected waves, which may finally destructively interfere. If we consider an infinite system 
with an infinitely large propagation, the transmission profile should become a delta-like distribution around the critical value 
$\Delta_{c}$. However, for a finite real system, the propagation distance will be finite and the transition broader, with a 
maximum transmission around the critical geometry. Close to the critical condition, most of the waves may experience a large 
transmission value through each diamond plaquette, and the collective behavior may induce an increasing transmitted front, which 
may reduce away from the optimal. We start by studying a lattice with 9 diamonds, as the fabricated one shown in Fig.~\ref{fig4}(a1), 
where we added several extra sites to the left and to the right such that we can differentiate the transmitted and reflected 
fronts. For this geometry, we numerically compute the transmitted beam after exciting the right edge and the lattice bulk, and 
show our results in Fig.~\ref{fig4}(a2) in red and green colors, respectively. Then, we numerically simulate lattices having $9$ 
randomly distributed plaquettes, for a left-edge excitation only, and obtain the averaged transmission over 100 different 
realizations [black curve in Fig.~\ref{fig4}(a2)]. We observe an excellent agreement of the obtained transmission for an 
individual realization and for the averaged of many~\cite{SM}, confirming our approach of implementing the experiments by 
varying the plaquette geometry only.

Fig.~\ref{fig4}(b1) shows the extracted data from the experiments performed on the lattice shown in Fig.~\ref{fig4}(a1), for 
different geometries $dx=dy$ and for $\lambda=750$ nm. Figs.~\ref{fig4}(b2) and (b3) show a compilation of the output images 
obtained after right-edge and bulk excitations [red and black data in Fig.~\ref{fig4}(b1)], respectively. We observe a clear 
increasing transmission of waves around the critical geometry $dx_{c}$, with a clear reduction of the reflected waves, as for 
example Fig.~\ref{fig4}(b2) at $dx=13\ \mu$m. These results show quite clearly the effect of the plaquette's geometry on the 
transport through a random DD lattice. Data and images show the strong consequences on transport for lattices away from the 
critical condition, for which randomly scattered waves interfere destructively, dynamically generating localization and reduced 
wave penetration for bulk and edge excitations. Independently of the random realization, a transition from insulating to 
conducting states is determined by the diamond geometry only. We observe an almost null transport of energy for $dx>16\ \mu$m 
and around $80\%$ of transmission for $dx\sim13\ \mu$m. It is indeed very remarkable that the transmission of energy on a 
lattice having several diamonds achieve a value around $80\%$, considering that for one plaquette, we obtained $\sim 90\%$.

Now, we emphasize the effect of the geometry on larger lattices, for example considering 16 random diamond plaquettes. 
Figs.~\ref{fig4}(c1) and (c2) show the intensity output profiles for $dx=13$ and $20\ \mu$m, respectively, for right-edge (top) 
and bulk (bottom) excitations. The contrast is more than evident when comparing only these two examples, which correspond exactly 
to the same random realization but just having different plaquette geometries. We observe clear dissemination of the energy for
the lattice at the critical geometry, with long penetration of waves from the edge and high dispersion from the bulk, while almost 
zero transport for $dx=20\ \mu$m (the black background indicates zero dispersed energy). We finally test another realization having 
$32$ random diamonds, see Figs.~\ref{fig4}(d). We observe clear bulk transport at $dx=dx_c$ and Anderson-like localization at $dx=20\ \mu$m. 

Finally, we test the effect of a flux $\Phi=\pi$ on each plaquette, obtained by inserting $P$ waveguides as it is sketched 
in Fig.~\ref{fig4}(e). In this way, we induce a negative coupling in the lower part of every diamond~\cite{MultiorbitalVic}, 
inducing destructive interference on any wave-packet passing through the lattice. Fig.~\ref{fig4}(f) shows three different 
lattices after bulk and edge excitations, for the indicated geometries. We notice that no transport of energy occurs in the 
lattice, independently of the plaquette geometry. The AB effect dramatically affects the transport properties of the system, 
and the conduction is simply forbidden by destructive interference. In this case, the localization occurs almost instantaneously 
in comparison to the localization process occurring in Anderson localization~\cite{Anderson-prb-1958}. 

In conclusion, we have shown an unexpected and rare phenomenon of complete delocalization of all single-particle states, for 
the entire band of allowed energies in a completely random disordered lattice model. We have demonstrated an exact analytical 
explanation behind the appearance of such an exceptional phenomenon, duly corroborated by comprehensive numerical simulations 
and direct experimental observation of this exceptional phenomenon for the light waves in completely disordered photonic lattices. 
Additionally, we have shown that our model also exhibits the Aharonov-Bohm caging phenomenon, giving rise to zero transmission 
of the waves for the same lattice geometries, when we turn on an effective $\pi$ flux in the system. Our results can be useful 
for potential applications in controlled optical transmission in a disordered medium, as well as in any wave physical system 
dealing with transport and localization phenomena.

\textit{Acknowledgements}.--- BP would like to thank Nagaland University for providing partial 
support through a start-up research grant for young faculties. BP also acknowledges Prof. Arunava 
Chakrabarti for simulating discussions on related projects done earlier. This research was supported 
in part by Millennium Science Initiative Program ICN17$\_$012 and ANID FONDECYT Grant 1231313.
%
\bibliography{refs}
\end{document}